\documentclass[a4paper, twocolumn]{article}
\usepackage[numbers]{natbib}
\usepackage[font=sf, labelfont={sf,bf}, margin=1cm]{caption}
\usepackage{amsmath}
\usepackage{amsfonts}
\usepackage{amsthm}
\usepackage{graphicx}
\usepackage{lscape}
\usepackage{anyfontsize}

\newcommand{\beginsupplement}{%
        \newpage
        \section*{Supplementary Materials}
        \setcounter{figure}{0}
        \renewcommand{\thefigure}{S\arabic{figure}}%
     }

\DeclareMathOperator*{\argmax}{arg\,max}

\usepackage{url}

\begin{document}

\title{Using genotype abundance to improve phylogenetic inference}

\author{William S. DeWitt III$^{1,2}$, Luka Mesin$^3$, Gabriel D. Victora$^3$,\\Vladimir N. Minin$^{4\ast}$ \& Frederick A. Matsen IV$^{1\ast}$}

\date{$^{1}$\small Computational Biology Program, Fred Hutchinson Cancer Research Center, Seattle, WA, USA\\
$^{2}$\small Department of Genome Sciences, University of Washington, Seattle, WA, USA\\
$^{3}$\small Laboratory of Lymphocyte Dynamics, The Rockefeller University, New York, NY, USA\\
$^{4}$\small Department of Statistics, University of California, Irvine, CA, USA\\
$\ast$ \footnotesize corresponding authors: VNM vminin@uci.edu, FAM matsen@fredhutch.org
}

\maketitle

\renewcommand{\abstractname}{\vspace{-\baselineskip}}
\abstract{\bf Modern biological techniques enable very dense genetic sampling of unfolding evolutionary histories, and thus frequently sample some genotypes multiple times.
This motivates strategies to incorporate genotype abundance information in phylogenetic inference.
In this paper, we synthesize a stochastic process model with standard sequence-based phylogenetic optimality, and show that tree estimation is substantially improved by doing so.
Our method is validated with extensive simulations and an experimental single-cell lineage tracing study of germinal center B cell receptor affinity maturation.
}

\section*{Introduction}
\label{sec:Intro}
Although phylogenetic inference methods were originally designed to elucidate the relationships between groups of organisms separated by eons of diversification, the last several decades have seen new phylogenetic methods for populations that are evolving on the timescale of experimental sampling \cite{Drummond2003-qx}.
This development is being spurred by new experimental techniques that enable deep sequencing at single-cell resolution, some of which enable quantification of original abundance.
For bulk sequencing, random barcodes can be used to quantify PCR template abundance \cite{Kivioja2011-kv,Jabara2011-ir,Brodin2015-rf}.
More recently, cell isolation \cite{Shapiro2013-ri} or combinatorial techniques \cite{Cusanovich2015-qj,dewitt2016,Howie2015-vp} have provided sequence data at single-cell resolution.
With such data, a given unique genotype---among many in the data---is represented in a measured number of cells.
The \emph{abundance} of a genotype can be read out as the number of cells bearing that genotype.
Here we demonstrate that incorporating genotype abundance improves phylogenetic inference for densely sampled evolutionary processes in which it is common to sample genotypes more than once.

We are motivated by the setting of B cell development in germinal centers.
B cells are the cells that make antibodies, or more generally \emph{immunoglobulins}.
Immunoglobulins are encoded by genes that undergo a stage of rapid Darwinian mutation and selection called \emph{affinity maturation} \cite{Mesin2016-ew}.
During affinity maturation, immunoglobulin is in its membrane-bound form, known as the \emph{B cell receptor} (BCR).
The biological function of this process is to develop BCRs with high-affinity for a pathogen-associated \emph{antigen} molecule, and later excrete large quantities of the associated antibody.

This affinity maturation process occurs in specialized sites called \emph{germinal centers} in lymph nodes, which have specific cellular organization to enable B cells to compete among each other to bind a specific antigen (proliferating more readily if they do) while mutating their BCRs via a mechanism called \emph{somatic hypermutation} (SHM).
Using micro-dissection, researchers can extract germinal centers from model animals and sequence the genes encoding their BCR directly \cite{Tas2016-lq, Kuraoka2016-zs}.
Lymph node samples are also available through autopsy \cite{Stern2014-ph} or fine needle aspirates from living subjects \cite{Havenar-Daughton2016-vk}.
Such samples provide a remarkable perspective on an ongoing evolutionary process.

Indeed, these samples contain a population of cells with BCRs that differentiated via SHM at various times and have various cellular abundances.
Because the natural selection process in germinal centers appears permissive to a variety of BCR-antigen affinities \cite{Tas2016-lq,Kuraoka2016-zs}, earlier-appearing BCRs are present at the same time as later-appearing BCRs.
The collection of descendants from a single founder cell in this process naturally form a phylogenetic tree.
However, it is a tree in which each genotype is associated with a given abundance, and such that older ancestral genotypes are present along with more recent appearances.
Reconstruction of phylogenetic trees from BCR data may benefit from methods designed to account for these distinctive features.

Standard sequence-based methods for inferring phylogenies fall into several classes according to their optimality criteria.
\emph{Maximum likelihood} methods posit a probabilistic substitution model on a phylogeny and find the tree that maximizes the probability of the observed data under this model \cite{felsenstein73, felsenstein81, felsenstein03}.
\emph{Bayesian} methods augment likelihood with a prior distribution over trees, branch lengths, and substitution model parameters, and approximate the posterior distribution of all the above variables by Markov chain Monte Carlo (MCMC) \cite{Huelsenbeck01, drummond15}.
\emph{Maximum parsimony} methods use combinatorial optimization to find the tree that minimizes the number of evolutionary events \cite{Eck66, Kluge69, Fitch71}.
Parsimony methods often result in degenerate inference, in which multiple trees achieve the same minimal number of events (i.e.\ maximum parsimony) \cite{Maddison1991-tc}.
Additional approaches include \emph{distance matrix} methods, which summarize the data by the distances between sequence pairs, and \emph{phylogenetic invariants}, which select topologies based on the value of polynomials calculated on character state pattern frequencies.
None of the above methods incorporate genotype abundance information, and it is standard for data with duplicated genotypes to be reduced to a list of \emph{deduplicated} unique genotypes before a phylogeny is inferred.

In this paper we show that genotype abundance is a rich source of information that can be productively integrated into phylogenetic inference, and we provide an open-source implementation to do so.
We incorporate abundance via a stochastic branching process with infinitely many types for which likelihoods are tractable, and show that it can be used to resolve degeneracy in parsimony-based optimality.
We first validate the procedure against simulations of germinal center BCR diversification.
We also empirically validate our method using an experimental lineage tracing approach combining multiphoton microscopy and single cell BCR sequencing, allowing us to study individual germinal center B cell lineages from brainbow mice.
Beyond the setting of BCR development, we foresee direct application to tumor phylogenetics in single-cell studies of cancer evolution (reviewed by Schwartz et al. \cite{Schwartz2017}), and single-cell implementations of lineage tracing based on genome editing technology \cite{McKennaaaf7907}.

\section*{New Approaches}

\subsection*{\bf Genotype-collapsed trees}

\begin{figure*}
  \centering
  \includegraphics[width=\linewidth]{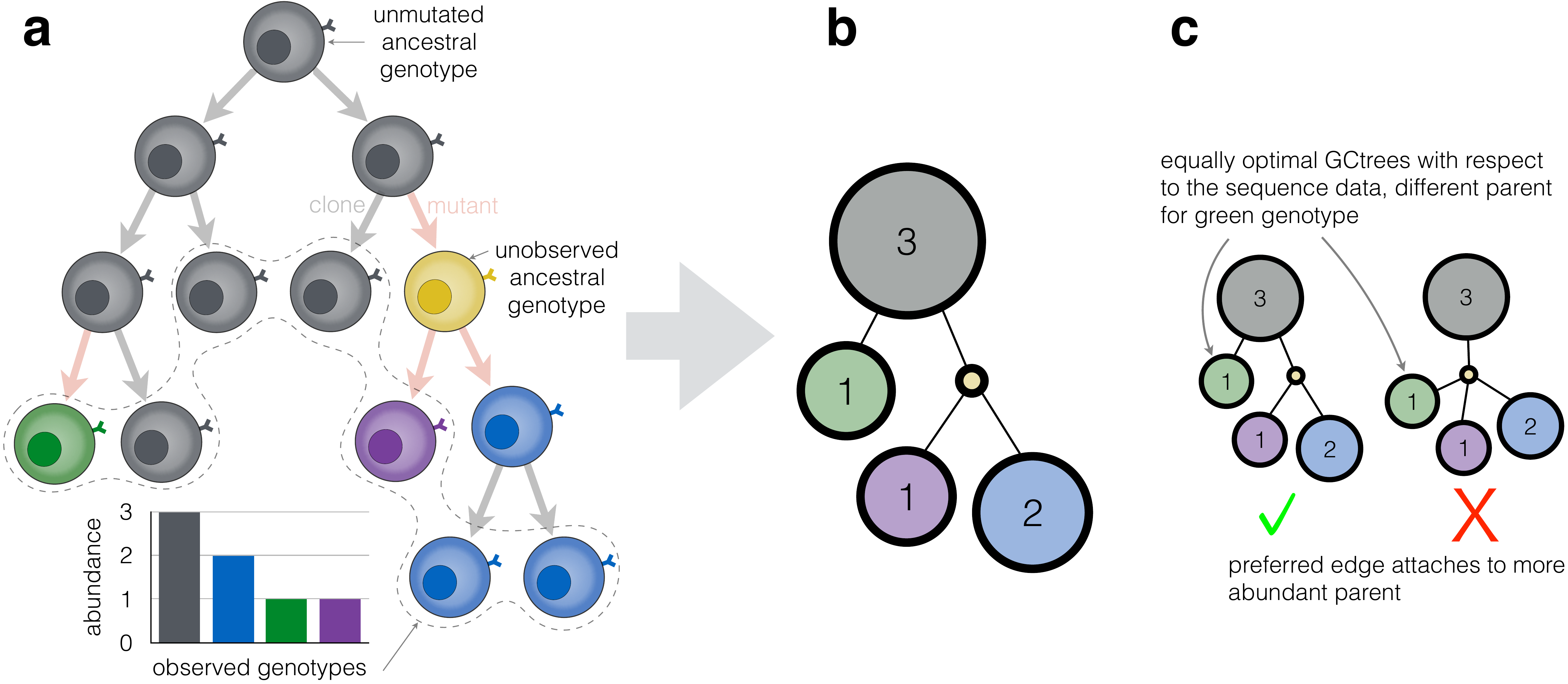}
  \vspace{-10pt}
  \caption{
  Genotype-collapsed trees.
  \textbf{(a.)} A diversifying B cell lineage is illustrated with distinct BCR genotypes colored.
  The final observed cells (enclosed by a dashed path) consist of genotypes at various abundances; note the yellow genotype is not observed.
  \textbf{(b.)} The corresponding genotype-collapsed tree (GCtree) describes the descent of distinct genotypes, and is our inferential goal.
  \textbf{(c.)} Genotype abundance informs topology inference.
  Two hypothetical GCtrees, equally optimal with respect to the sequence data, propose two possible parents of the green genotype---the gray and yellow genotypes (the yellow genotype was not sampled and thus has a small circle with no number inside).
  Intuitively, the abundance information indicates that the tree on the left is preferable because the more abundant parent is more likely to have generated mutant descendants.}
  \label{fig:gctree}
\end{figure*}

Given sequence data obtained from a diversifying cellular \emph{lineage tree} (Figure~\ref{fig:gctree}a), our goal is to infer the \emph{genotype-collapsed tree} (GCtree) defining the lineage of distinct genotypes and their observed abundances (Figure~\ref{fig:gctree}b).
The GCtree is constructed from the lineage tree by collapsing subtrees composed of cells with identical genotype to a single node annotated with its final cellular abundance.
Our data consists of the genotypes sampled at least once in the GCtree, along with their associated abundances.
Under the \emph{infinite types} assumption that every mutant daughter generates a novel genotype, each genotype can be identified with one subtree in the original lineage tree.
We are not claiming any originality in the GCtree definition, but it is useful to have a word for this object.

We note that, unlike standard phylogenetic trees where only leaf nodes represent observed genotypes, GCtree internal nodes represent observed genotypes if they are annotated with non-zero abundance.
Although not leaves \emph{per se} in the GCtree, a nonzero abundance represents a clonal sub-lineage that resulted in a nonzero number of leaves of that genotype in the lineage tree.
A node in the GCtree, along with its descending edges, summarizes the lineage outcome for a given genotype as its mutant offspring clades and the number of its clonal leaves.
Because this summary does not completely specify the genotype's clonal lineage structure (Figure~\ref{fig:model}c), several branching structures may be consistent with a given node, and we have no information with which to distinguish between the various lineage trees consistent with a GCtree.
Hence, our goal is to infer the GCtree topology.

\subsection*{\bf Parsimony with a prior}

BCR sequence data from a germinal center sample has the following characteristics from the perspective of phylogenetics: genotypes have abundances, there is a limited amount of mutation between genotypes, and ancestral genotypes are present along with later ones.
The latter two features suggest maximum parsimony as a useful tool because of the limited amount of mutation and because ancestral genotypes can be assigned to internal nodes of the tree (although recent Bayesian methods can do such assignment as well \cite{Gavryushkina2014-qp,Gavryushkina2015-ew}).
For these reasons, parsimony has been used extensively in B cell sequence analysis \cite{Barak2008-fw,Stern2014-ph}.
Because having many duplicate sequences inhibits efficient tree space traversal, these studies have inferred trees using the unique genotypes (BCR sequences).
This ignores the varying cellular abundances of the observed genotypes.

Here we wish to use a branching process model to rank trees that are equally optimal according to sequence-level optimality criteria.
Indeed, maximum parsimony often results in degenerate inference: there are many trees that are maximally optimal \cite{Maddison1991-tc}.
We refer to these trees as a \emph{parsimony forest}.
In later sections we show, using \textit{in silico} and empirical data, that parsimony degeneracy is common and often severe for BCR sequencing data, and that parsimony forests exhibit substantial variation in phylogenetic accuracy.
It is common practice to arbitrarily select one tree in the parsimony forest at random, without regard for this variability in inference accuracy.
Instead, we will rank trees in the parsimony forest with an auxiliary likelihood that incorporates abundance information, thereby resolving this degeneracy.

Genotype abundance is an additional source of information for phylogenetics, using the simple intuition that more abundant genotypes are more likely to have more mutant descendant genotypes.
This intuition makes sense because relative sample abundance is a reasonable estimator of relative total historical abundance, and total historical abundance is related to the number of mutant offspring---i.e. genotypes with larger abundance are likely to have more mutant descendant genotypes simply because there are more individuals available to mutate.
The number of mutant offspring genotypes is in turn related to the number of surviving mutant offspring sampled.
Thus, given two equally parsimonious trees, this intuition would prefer the tree that has more mutant descendants of a frequently observed node (Figure~\ref{fig:gctree}c).
We formalize this intuition using a stochastic process model for the phylogenetic development of germinal centers, and integrate this model with sequence-based tree optimality via empirical Bayes.

In this stochastic process model, a GCtree node $i$ has a random number $T_i \in\mathbb N$ of mutant children (i.e.\ descending edges) and a random abundance $A_i\in\mathbb N$.
We will index nodes in a ``level order'' as follows, which is well defined given an embedding of the tree into the plane.
Index $1$ refers to the root node, and $2$ through $1+T_1$ refer to the children of the root node.
The level-order continues in order through all tree nodes of the same level before nodes at the next level.
Adopting this level-ordering convention, a GCtree containing $N$ nodes is specified by integer-valued random vectors giving the (planar) topology $\mathbf{T} = (T_1,\dots, T_N)$, and abundances $\mathbf{A} = (A_1,\dots, A_N)$.
We also have the observed genotype sequences associated with each node $\mathbf{G}=(G_1,\dots,G_N)$.

\begin{figure*}
\centering
\includegraphics[width=\linewidth]{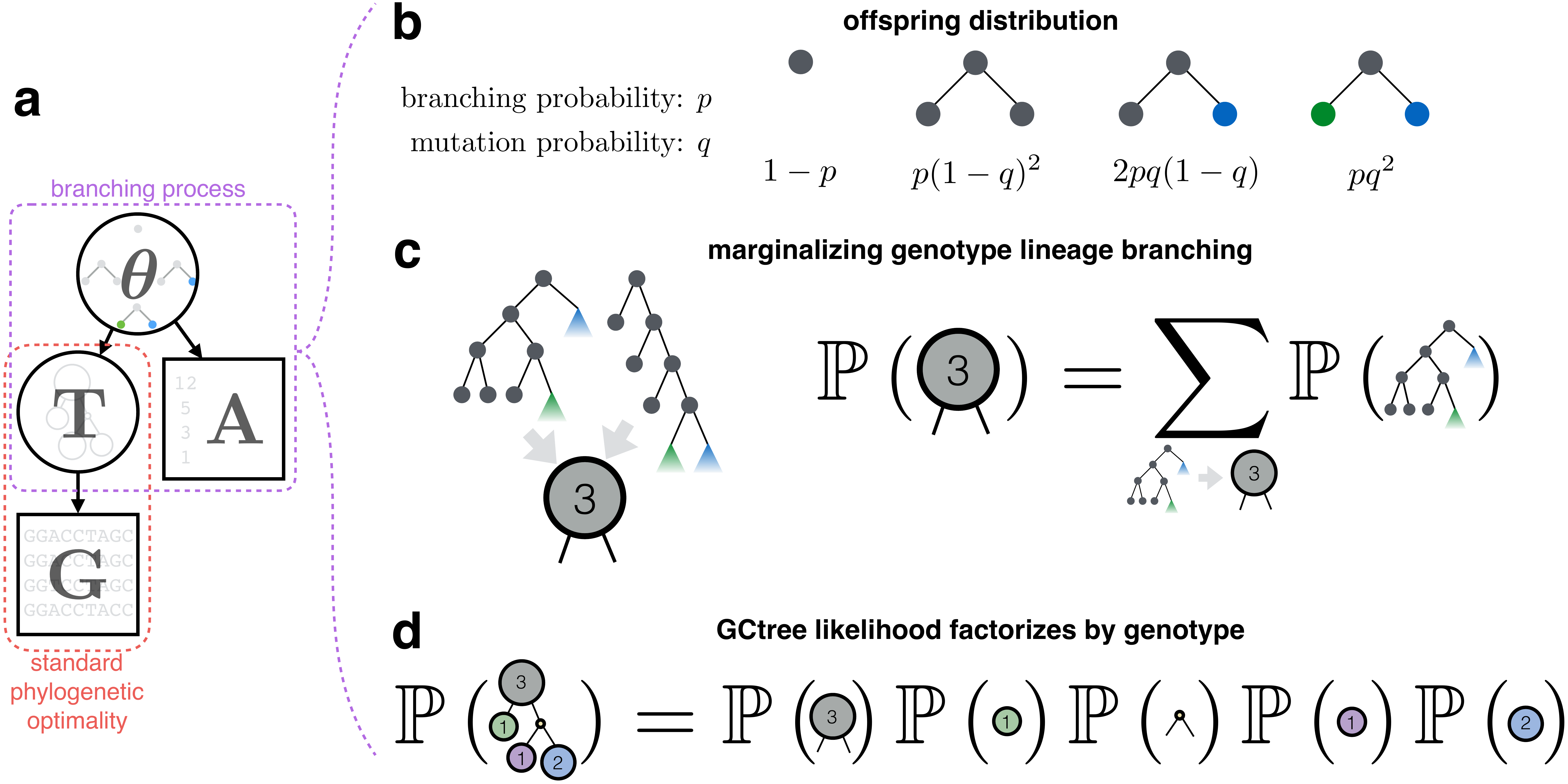}
 \vspace{-10pt}
\caption{
Modeling sequences equipped with abundances.
\textbf{(a.)} Both genotype sequence data $\mathbf{G}$ and genotype abundance data $\mathbf{A}$ inform tree topology $\mathbf{T}$.
As illustrated in this probabilistic graphical model, we assume independence between $\mathbf{G}$ and $\mathbf{A}$ conditioned on $\mathbf{T}$ rather than a fully joint model of $\mathbf{G}$, $\mathbf{A}$, and $\mathbf{T}$.
This facilitates using standard sequence-based phylogenetic optimality for $\mathbf{G}$, augmented with a branching process (with parameters $\boldsymbol\theta$) for $\mathbf{A}$.
\textbf{(b.)} For the binary infinite-type Galton-Watson process, $\boldsymbol\theta = (p, q)$.
Four possible branching events characterize the offspring distribution common to all nodes.
A node may bifurcate (with probability $p$) or terminate, and upon bifurcating its descendants each may be a mutant (with probability $q$).
\textbf{(c.)} A GCtree node specifies a genotype's clonal leaf count and number of descendant genotypes, but not lineage details.
The likelihood of a GCtree node marginalizes over consistent lineage branching outcomes.
\textbf{(d.)} GCtree likelihood factorizes into the product of likelihoods for each genotype.}
\label{fig:model}
\end{figure*}

A complete diversification model would give a joint distribution on $\mathbf{T}$, $\mathbf{G}$, and $\mathbf{A}$.
As an approximation to such a model, facilitating use of existing sequence-based optimality methods, we propose a generative model containing conditional independences as follows (Figure~\ref{fig:model}a).
First, we model abundances $\mathbf{A}$ and tree topology $\mathbf{T}$ as being drawn from a branching process likelihood, conditioned on parameters $\boldsymbol\theta$ (characterizing birth, death, and mutation rates in the underlying lineage tree): $\mathbb{P}\left(\mathbf{A}, \mathbf{T}\mid \boldsymbol\theta\right)$.
This stochastic process likelihood will capture the intuition (described above) that more abundant genotypes are likely to have more mutant descendant genotypes.
Next, we assume that genotype sequences $\mathbf{G}$ are generated by a mutation model conditioned on the fixed tree $\mathbf{T}$, independent of $\mathbf{A}$.
This sequence-based optimality is captured by a distribution over $\mathbf{G}$ dependent only on $\mathbf{T}$: $\mathbb{P}\left(\mathbf{G}\mid \mathbf{T}\right)$.
The lack of direct dependence of $\mathbf{G}$ on $\mathbf{A}$ constitutes an approximation to a more realistic sequence-valued branching process.
However, this formulation has the advantage that it allows us to leverage standard sequence-based phylogenetic optimality in the specification of $\mathbb{P}\left(\mathbf{G}\mid \mathbf{T}\right)$.
In a later section (\textit{In silico} validation), we validate this approximation with simulations that do not assume this conditional independence.

In an empirical Bayes treatment (see Materials and Methods for details), a maximum likelihood estimate for the branching process parameters, $\hat{\boldsymbol{\theta}}$, can be obtained by marginalizing $\mathbf{T}$, and this in turn can be used to approximate a posterior over $\mathbf{T}$ conditioned on the data $\mathbf{G}$ and $\mathbf{A}$ (as well as $\hat{\boldsymbol{\theta}}$).
Using parsimony as our sequence-based optimality, one can rank trees in the parsimony forest (denoted $\mathcal{T}_\mathbf{G}$) according to the GCtree likelihood.
We encode the parsimony criteria in $\mathbb{P}\left(\mathbf{G}\mid \mathbf{T}\right)$ by assigning uniform weight to the trees in $\mathcal{T}_\mathbf{G}$, and zero to the other trees.
This gives the following approximate maximum a posteriori tree:
\begin{equation}
\hat{\mathbf{T}} = \argmax_{\mathbf{T}\in\mathcal{T}_\mathbf{G}} \mathbb{P}\left(\mathbf{A}, \mathbf{T}\mid \hat{\boldsymbol{\theta}}\right),
\label{eqn:MLE}
\end{equation}
where the point estimate $\hat{\boldsymbol{\theta}}$ is given by
\begin{equation}
\hat{\boldsymbol{\theta}} = \argmax_{\boldsymbol{\theta}} \sum_{\mathbf{T}\in\mathcal{T}_\mathbf{G}}\mathbb{P}\left(\mathbf{A}, \mathbf{T}\mid\boldsymbol\theta\right).
\label{eqn:theta}
\end{equation}
Next we turn to explicitly defining the GCtree likelihood $\mathbb{P}\left(\mathbf{A}, \mathbf{T}\mid\boldsymbol\theta\right)$.

\subsection*{\bf A stochastic process model of abundance}

To compute likelihoods $\mathbb{P}\left(\mathbf{A}, \mathbf{T}\mid\boldsymbol\theta\right)$ for GCtrees (Figure~\ref{fig:gctree}b), we model the lineage tree (Figure~\ref{fig:gctree}a) as a subcritical infinite-type binary Galton-Watson (branching) process \cite{harris2002theory} in which extinct leaf nodes correspond to observed cells.
All mutations in an infinite-type process result in a novel genotype, embodying the assumption that each genotype can be identified with one subtree.
Subcriticality ensures that the branching process terminates in finite time, so an explicit sampling time is not needed.
The process is initiated with a single cell (a naive germinal center B cell before affinity maturation ensues), and runs to eventual extinction.
This model is highly idealized and unable to capture many biological realisms of B cell affinity maturation and the sampling process.
However, as we show in our validations, it is useful as a minimal model for leveraging genotype abundance information in a tractable likelihood.

The offspring distribution for our process, governing reproduction and mutation for all lineage tree nodes at all time steps, is specified by two parameters: the binary branching probability $p$, and the mutation probability $q$.
Because the offspring distribution is independent of type, subcriticality simply requires that the expected number of offspring of any node is less than 1, in this case equivalent to $p < 0.5$.
In this case a ``mutation'' is an event that causes the evolving lineage to change to a novel genotype (under the infinite-types assumption).
Thus the corresponding offspring distribution supports four distinct branching events (Figure~\ref{fig:model}b).
Letting $C$ and $M$ denote the (random) number of clonal and mutant offspring of any given node in the lineage tree, respectively, the offspring distribution is
\begin{equation}
    \mathbb P\left(C=c, M=m\right) =
    \begin{cases}
          1-p & c=m=0,\\
          p(1-q)^2 & c=2, m=0,\\
          2pq(1-q) & c=m=1,\\
          pq^2 & c=0, m=2,\\
          0& \text{otherwise}.
    \end{cases}
  \label{eqn:offspring}
\end{equation}

We can compute the likelihood of a hypothetical binary lineage tree simply by evaluating \eqref{eqn:offspring} at each node in the tree and multiplying the results.
The likelihood for a GCtree is then given by summing over all possible binary lineage trees that are consistent with that GCtree (i.e.\ that give the same GCtree when collapsing by genotype), thus marginalizing out the details of intra-genotype branching events that give rise to the same abundance.
Here we show how to calculate the GCtree likelihood directly for the simple offspring distribution \eqref{eqn:offspring}.
Other work \cite{Bertoin2009-jy} has described how to calculate statistics of the infinite-type branching process with a general subcritical offspring distribution.

First consider the likelihood for an individual node in the GCtree, say the root node, in the context of the branching process described above.
A GCtree node $i$ is specified by its abundance $A_i$ and the number of edges descending from it $T_i$ (both random variables).
There are, in general, multiple distinct branching process realizations for genotype $i$ that result in $A_i=a$ clonal leaves and $T_i=\tau$ mutations off the genotype $i$ lineage subtree (Figure~\ref{fig:model}c).
Determining the likelihood of node $i$ in the GCtree under this process, which we denote by $f_{a \tau}(p, q) = \mathbb{P}\left(A_i=a,T_i=\tau\mid \boldsymbol\theta=(p, q)\right)$, requires marginalizing over all such genotype lineage subtrees.
In Materials and Methods we derive a recurrence for $f_{a \tau}(p, q)$ by marginalizing over the outcome of the branching event at the root of the lineage subtree for genotype $i$, and show that the GCtree node likelihood $f_{a \tau}(p, q)$ can be computed by dynamic programming.

A complete GCtree containing $N$ nodes is specified by level-ordering the nodes as described above $\mathbf{T} = (T_1,\dots, T_N)$, $\mathbf{A} = (A_1,\dots, A_N)$.
Because the same offspring distribution generates the lineage branching of each genotype subtree, the same recurrence can be applied to all GCtree nodes.
Specifically, we show in Materials and Methods that the joint distribution over all nodes in a GCtree factorizes by genotype (Figure~\ref{fig:model}d):
\begin{equation}
\begin{aligned}
\mathbb{P}\left(\mathbf{T}=(\tau_1,\dots,\tau_N), \mathbf{A}=(a_1,\dots,a_N)\mid\boldsymbol\theta=(p, q)\right)
\\
= \prod_{i=1}^N f_{a_i \tau_i}\left(p, q\right).
\end{aligned}
\label{eqn:likelihood}
\end{equation}

\begin{figure}[h]
\centering
\includegraphics[width=.5\linewidth]{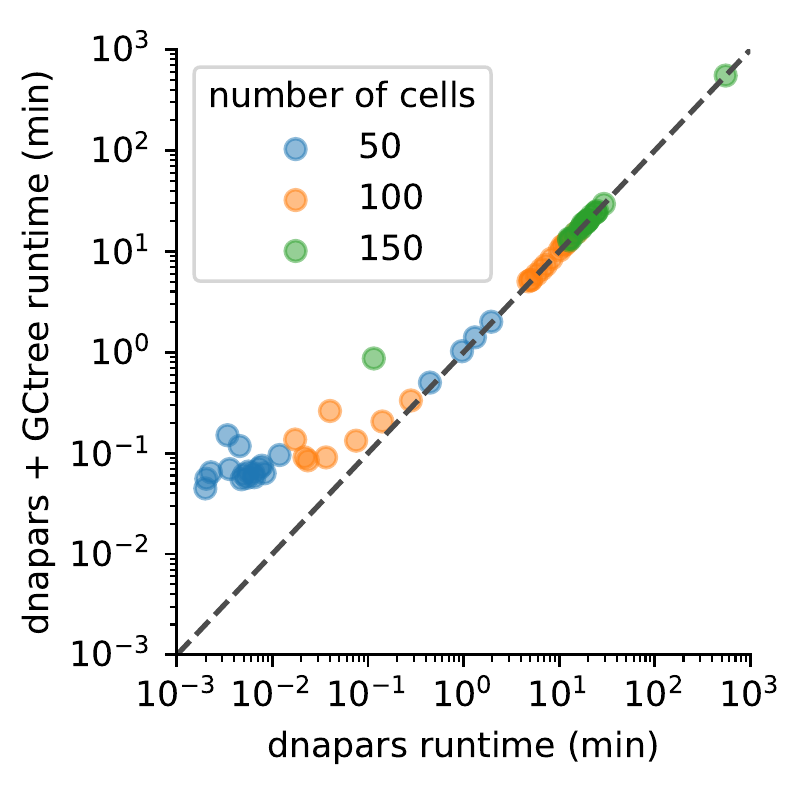}
\caption{
Runtime experiments.
Runtime for generating parsimony trees with \texttt{dnapars} and ranking using \texttt{GCtree} are shown.
Fixed simulation parameters were $\lambda=1.5$, $\lambda_0=.25$, and 20 simulations were performed at each value for the number of cells ($N=50$, $N=100$, $N=150$).
These runtime experiments were performed on a laptop with a 2.9GHz CPU and 16GB RAM.}
\label{fig:runtime}
\end{figure}

Using dynamic programming and factorization by genotype, the computational complexity of the GCtree likelihood is $\mathcal{O}(\max(A)\max(T)+N)$.
Ranking parsimony trees with \texttt{GCtree} requires a polynomial increase in runtime compared with finding the parsimony forest, which is itself NP-hard \cite{FOULDS198243}.
Figure~\ref{fig:runtime} depicts runtime from simulations of various size, and shows that, in practice, this increased runtime is negligible.

A computational implementation of the inference method above is available at \url{http://github.com/matsengrp/gctree}.
The \texttt{GCtree} inference subprogram accepts sequence data in \texttt{FASTA} or \texttt{PHYLIP} format, determines a parsimony forest from the unique sequences using the \texttt{dnapars} program from the \texttt{PHYLIP} package \cite{Felsenstein2005}, determines the genotype-collapsed form of these trees and outputs tree visualizations using the \texttt{ETE} package \cite{doi:10.1093/molbev/msw046}, and ranks them according to their GCtree likelihood using the sequence abundances.
Bootstrap analysis is also implemented, providing confidence values of each split in the maximum likelihood GCtree.
The GCtree maximizing the branching process likelihood (with optional bootstrap support) is the inference result.
Next we show that resolving parsimony degeneracy using \texttt{GCtree} substantially increases both accuracy and precision of phylogenetic inference.

\section*{Results}

\subsection*{\bf \textit{In silico} validation}

To explore the accuracy and robustness of \texttt{GCtree} inference, we developed a simulation subprogram to generate random lineages starting with a naive BCR sequence.
For simulated lineages, true trees can be compared against those inferred with the \texttt{GCtree} inference subprogram.
The stochastic process model used in \texttt{GCtree} inference is intended as a minimal model (in terms of biological realism) that captures the intuition that genotype abundance is relevant to phylogenetic reconstruction.
Experimental data need not obey our simplifying assumptions, thus we set out to test \texttt{GCtree}'s robustness to deviations of the data generating process from the inferential model.

A simulation process was implemented that includes biological realisms of B cells undergoing SHM (and violates inferential assumptions).
These realisms of simulation---detailed in Materials and Methods---include: branching process multifurcations (controlled by a parameter $\lambda$, the expected number of children of a node in the cell lineage tree), sequence context sensitive mutations \cite{Dunn-Walters2360, Spencer5170} (with a baseline-line mutation rate $\lambda_0$, and a context-specific mutational model with 5mer mutabilities taken from \cite{Yaari2013-uw}), explicit sampling time ($t$, or $N$ representing the number of cells desired in the sampled generation), incomplete sampling (the number of cells to sample $n \leq N$), and repeated genotypes allowed (deviation from the infinite-type assumption).
This constitutes a more challenging validation than simply simulating under the same assumptions that had been invoked for tractability of the inferential framework.

\begin{figure*}
\centering
\includegraphics[width=\linewidth]{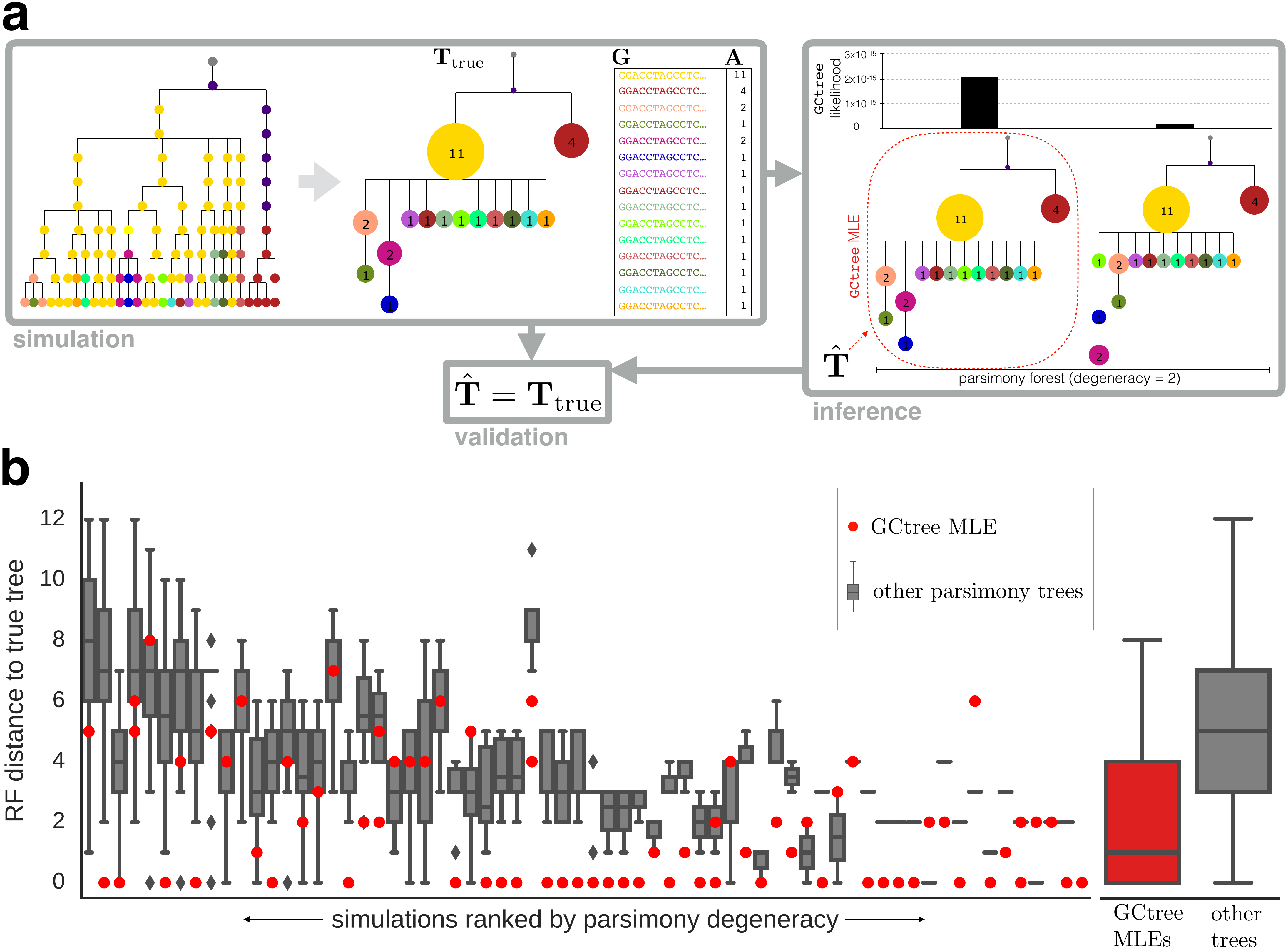}
 \vspace{-10pt}
\caption{
\textit{In silico} validation of \texttt{GCtree} inference.
\textbf{(a.)}
Demonstrating the simulation--inference--validation workflow, a small simulation resulted in two equally maximally parsimonious trees, and the one inferred using \texttt{GCtree} was correct.
The initial sequence was a naive BCR V gene from the experimental data described in Materials and Methods.
Branch lengths in the cell lineage tree (left) correspond to simulation time steps, while those in collapsed trees correspond to sequence edit distance.
\textbf{(b.)}
100 simulations were performed with parameters calibrated using the BCR sequencing data and summary statistics described in Materials and Methods.
Of 100 simulations, 66 resulted in parsimony degeneracy, with an average degeneracy of 12 and a maximum degeneracy of 124.
For each of these 66, we show the distribution of Robinson--Foulds (RF) distance of trees in the parsimony forest to the true tree.
``RF'' denotes a modified Robinson-Foulds distance: since nonzero abundance internal nodes in GCtrees represent observed taxa, RF distance was computed as if all such nodes had an additional descendant leaf representing that taxon.
GCtree MLEs (red) tend to be better reconstructions of the true tree than other parsimony trees (gray boxes).
Four simulations resulted in a tie for the GCtree MLE, and the two tied trees in these cases are both displayed in red.
Aggregated data across all simulations are depicted on the right, clearly indicating superior reconstructions from \texttt{GCtree}.}
\label{fig:simulation}
\end{figure*}

Our \textit{in silico} validation workflow is demonstrated in Figure~\ref{fig:simulation}a for a small simulation that resulted in a parsimony forest with just two equally parsimonious trees.
The output of the simulation software consists of \texttt{FASTA} data (sequences and their abundances), visualizations of the lineage tree and its GCtree equivalent, and a file containing the true GCtree structure.
The \texttt{GCtree} inference subprogram can then be run on the \texttt{FASTA} data, and the resulting inferred GCtree compared to the true GCtree (in this case they were identical).
To calibrate simulation parameters, we defined summary statistics on sequence data with abundance information, and tuned parameters to produce data similar to experimental BCR sequencing data under these statistics (see Materials and Methods).

Our validation shows that using abundance information via a branching process likelihood can substantially improve inference results (Figure~\ref{fig:simulation}b).
For each simulation we ranked otherwise degenerately optimal parsimony trees using \texttt{GCtree}.
For each parsimony forest, we compared the GCtrees in the forest to the true GCtree for that simulation using the Robinson--Foulds (RF) distance \cite{ROBINSON1981131} as a measure of tree reconstruction accuracy.
The maximum likelihood GCtree tends to be closer to the true tree than other equally parsimonious trees, which vary widely in accuracy, showing that GCtree is able to leverage abundance data to resolve parsimony degeneracy and improve the accuracy of tree reconstruction in this simulation regime.

\subsection*{\bf Empirical validation}

We next performed a biological validation by investigating if \texttt{GCtree} improves inference according to biological criteria using real germinal center BCR sequence data.
The BCR is a heterodimer encoded by the immunoglobulin heavy chain (IgH) and immunoglobulin light chain (IgL) loci.
Both loci undergo V(D)J recombination, and then evolve in tandem during affinity maturation.
By obtaining matched sequences from both loci using single-cell isolation, we have two independent data sets to inform the same phylogeny of distinct cells (each of which is associated with a single IgH sequence and single IgL sequence).
Performing separate and independent IgH and IgL tree inference, we can then validate \texttt{GCtree} by comparing the inferred IgH tree to the inferred IgL tree.
If the GCtree likelihood \eqref{eqn:likelihood} meaningfully ranks equally parsimonious trees, then the two MLE trees (IgH and IgL) would be expected to be more correct reconstructions than the other parsimony trees.
Thus, we are to expect that the two MLE trees are more similar to each other (in terms of the lineage of distinct cells) than other pairs of IgH and IgL parsimony trees (which, if they are more distorted phylogenies, should show less concordance in the partitioning of the distinct cells).
Conversely, if the GCtree likelihood is not meaningfully ranking trees, we expect that the MLE IgH and IgL trees will not be significantly closer to each other than other pairs of IgH and IgL parsimony trees.

\begin{figure*}
\centering
\includegraphics[width=.9\linewidth]{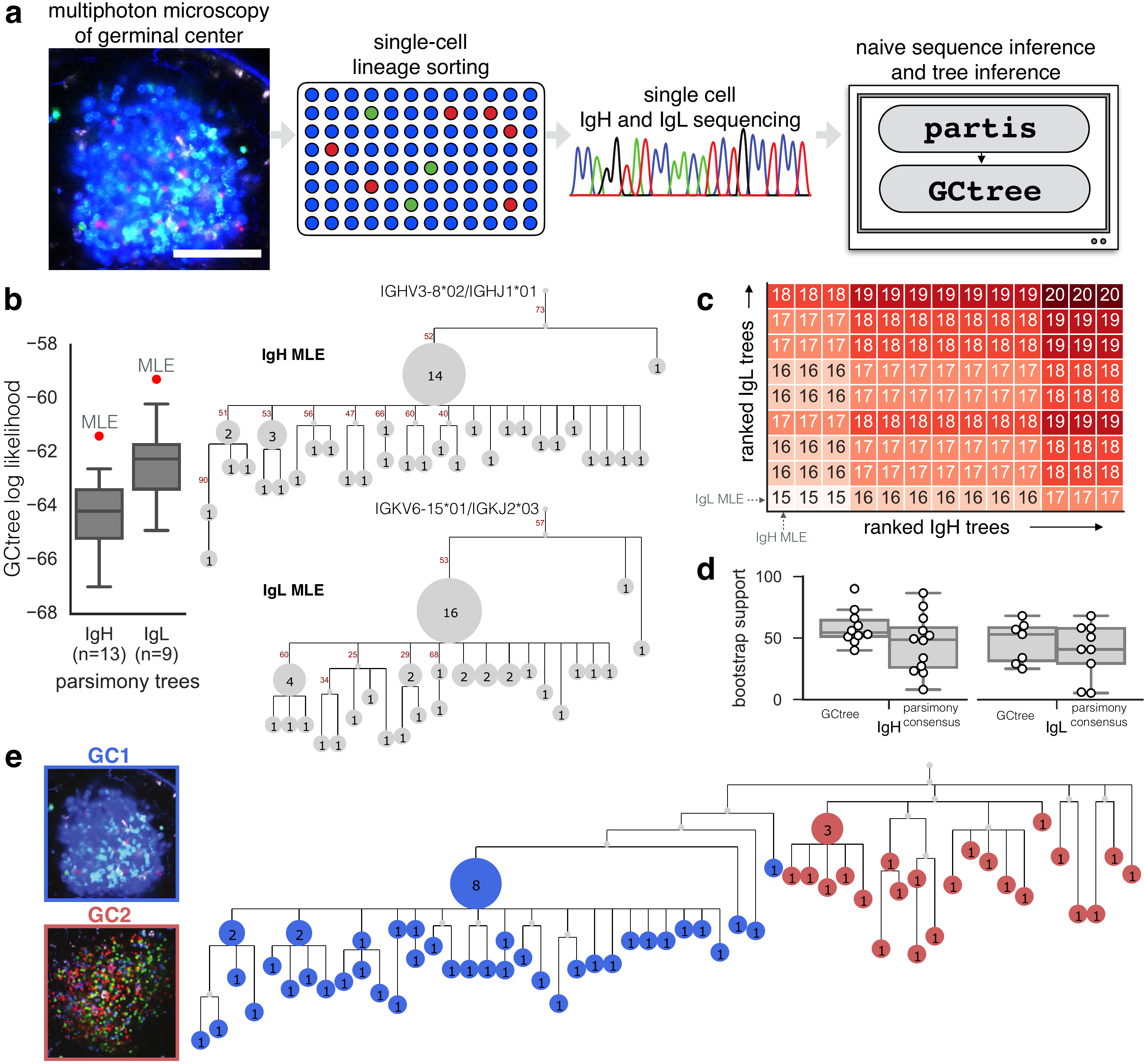}
 \vspace{10pt}
\caption{
Empirical validation using lineage tracing and single cell germinal center BCR sequencing.
\textbf{(a.)} A multiphoton image of a germinal center reveals a dominant blue lineage (scale bar 100$\mu$m).
This lineage was sorted, and 48 cells sequenced to determine IgH and IgL genotypes of each.
These sequences were analyzed with \texttt{partis} \cite{partis1,partis2} to infer naive (pre-affinity-maturation) ancestor sequences using germline genetic information, and trees were inferred with \texttt{GCtree}.
\textbf{(b.)} \texttt{GCtree} inference was performed separately for IgH and IgL loci, resulting in parsimony degeneracies of 13 and 9, respectively.
Maximum likelihood GCtrees for each locus are indicated in red and the GCtrees with annotated abundance are shown.
Roots are labeled with the gene annotations of the naive state inferred using \texttt{partis}.
Small unnumbered nodes indicate inferred unobserved ancestral genotypes.
Numbered edges indicate support in 100 bootstrap samples.
\textbf{(c.)} All possible pairings of IgH and IgL parsimony trees were compared in terms of the Robinson-Foulds distance between the IgH and IgL trees, labeled by cell identity.
IgH and IgL parsimony trees are ordered by GCtree likelihood rank in columns and rows, respectively.
Grid values show RF distance between each IgH/IgL pair.
MLE trees result in more consistent cell lineage reconstructions between IgH and IgL (smaller RF values).
\textbf{(d.)} For each locus, distributions of bootstrap support values are shown for the tree inferred by \texttt{GCtree} and for a majority rule consensus tree of all trees in the parsimony forest.
The latter contain more partitions with very low support.
\textbf{(e.)} Using additional data from a second germinal center from the same lymph node that had the same naive BCR sequence, \texttt{GCtree} correctly resolves the two germinal centers as distinct clades (as did other lower ranked parsimony trees).
}
\label{fig:empval}
\end{figure*}

We used data from a previously reported experiment in which multiphoton microscopy and BCR sequencing were combined to resolve individual germinal center B cell lineages from mouse lymph nodes 20 days after subcutaneous immunization with alum-adsorbed chicken gamma globulin \cite{Tas2016-lq} (see Materials and Methods).
\emph{Brainbow} mice were used for multicolor cell fate mapping, enabling B cells and their progeny to be permanently tagged with different fluorescent proteins.
In-situ photo-activation followed by fluorescence-activated cell sorting yielded B cells from a color-dominant germinal center (Figure~\ref{fig:empval}a, left).
BCR sequences were obtained for 48 cells in this lineage by single cell mRNA sequencing of the IgH and IgL loci, resulting in 32 distinct IgH and 26 distinct IgL genotypes due to SHM mutations acquired through affinity maturation.
The unmutated naive IgH and IgL V(D)J rearranged sequences (not observed) were inferred with \texttt{partis} using each set of 48 sequences (IgH and IgL) as a clonal family using germline genetic information \cite{partis1, partis2}.
These naive sequences were used as outgroups for rooting parsimony trees.

\texttt{GCtree} results are depicted in Figure~\ref{fig:empval}b.
Parsimony analysis resulted in degeneracy for both loci, with 13 equally parsimonious trees for IgH, and 9 for IgL.
Empirical Bayes point estimation according to \eqref{eqn:theta} yielded $\hat p = 0.495$, $\hat q = 0.388$ (IgH) and $\hat p = 0.495$, $\hat q = 0.304$ (IgL).
GCtree likelihoods \eqref{eqn:likelihood} were computed to rank the equally parsimonious trees, and the MLE trees are shown with support values among 100 bootstrap samples (see Materials and Methods).
Because the binary Galton-Watson process assigns probability zero to a GCtree node with frequency zero and one mutant descendant, the unobserved naive root node (which had one descendant after rerooting and collapsing identical genotypes in all parsimony trees) was given a unit pseudocount.

We then compared the concordance between pairs of heavy and light trees.
Since both IgH and IgL loci have been recorded from the same set of 48 cells, the units of cell abundance in an IgH GCtree map to the units of cell abundance from an IgL GCtree (i.e.\ each cell identity among the 48 is associated with an IgH genotype and an IgL genotype).
We can then consider the consistency of a given IgH tree and a given IgL tree in terms of the lineage of the 48 cell identities.
For each possible pairing of an IgH parsimony tree with a IgL parsimony tree, we computed the RF distance \cite{ROBINSON1981131} between the two trees using the cell identities (rather than the genotype sequences) to define splits.
We observed that the GCtree MLE based on IgH sequences and GCtree MLE based on IgL sequences form the most concordant pair among all pairs of parsimony trees (Figure~\ref{fig:empval}c).
Moreover, pairs of parsimony trees that contained at least one GCtree MLE tree ranked consistently higher in terms of their similarity.

We assessed confidence in \texttt{GCtree} partitions by comparing bootstrap support values in \texttt{GCtree} trees to those from the majority-rule consensus parsimony trees made using the \texttt{consense} program from the \texttt{PHYLIP} package \cite{Felsenstein2005}.
We observed the latter contained an excess of very low confidence partitions (Figure~\ref{fig:empval}d, Figure~\ref{fig:support}).
These results demonstrate that parsimony reconstructions for real BCR data sets suffer from degeneracy, and that GCtree likelihood can correctly resolve this degeneracy by incorporating abundance information ignored by previously published methods.

Finally, using data collected from a second germinal center from the same lymph node, we tested \texttt{GCtree}'s ability to correctly group cells from each germinal center into separate clades when run on combined data from both germinal centers.
The two germinal center sequence data sets appeared to have the same naive BCR sequence (IgH and IgL), indicating they were both seeded from the same B cell lineage.
Concatenating the IgH and IgL sequences for each cell in each germinal center, we used \texttt{GCtree} to infer a single tree for all cells from both germinal centers (Figure~\ref{fig:empval}e, Figure~\ref{fig:labelled_tree}).
\texttt{GCtree} correctly resolved the two germinal centers as distinct clades (we note that all the parsimony trees had this feature, regardless of likelihood rank).
This demonstrates the phylogenetic resolvability of germinal centers with the same naive BCR diversifying under selection for the same antigen specificity.

\section*{Discussion}

We have shown that genotype abundance information can be productively incorporated in phylogenetic inference.
By augmenting standard sequence-based phylogenetic optimality with a stochastic process likelihood, we were able to implement abundance-aware inference as a processing step downstream of results from an existing and widely used parsimony tree inference tool.
We have shown that our method---implemented in the publicly available \texttt{GCtree} package---is useful for inferring B cell receptor affinity maturation lineages.
Although branching processes have been used previously to infer parameters of BCR evolution \cite{Kleinstein2003-to, Magori-Cohen2006} and construct SHM lineage trees from error-prone bulk sequencing reads \cite{Sok2013-xl}, to our knowledge we are the first to use branching processes to sharpen phylogenetic inference for BCRs sequenced at single-cell resolution from germinal centers.

We believe \texttt{GCtree} will find use in other settings where sequence data from dense quantitative sampling of diversifying loci are available.
Studies of cancer evolution are increasingly performed with single-cell resolved sequencing, however most tumor phylogenetics approaches use standard phylogenetic methods (reviewed by Schwartz et al. \cite{Schwartz2017}) that do not model genotype abundance.
Exceptions include \texttt{OncoNEM} \cite{Ross2016-re} and \texttt{SCITE} \cite{Jahn2016-lt}, both of which leverage single-cell data for tumor phylogenetic inference that is robust to genotyping errors and missing data, but do not aim to capture the intuition that genotype abundance and the number of direct mutant descendants are related.
Single-cell implementations of lineage tracing based on genome editing technology \cite{McKennaaaf7907} may also benefit from reconstruction methods that model the abundance of observed editing target states, since cell types may vary widely in rates of proliferation.

Using parsimony as our sequence-based optimality resulted in particularly simple results, as the tree space necessary to explore is exactly the degenerate parsimony forest.
However, our empirical Bayes formulation is agnostic to the particular choice of sequence-based optimality, so in the future we envision augmenting likelihood-based sequence optimality.
This will require more computationally expensive tree space search and sampling schemes.

In contrast to \texttt{GCtree}, a fully Bayesian approach to incorporate genotype abundance could use the full set of sequences (without deduplication) in a Bayesian phylogenetics package---such as \texttt{BEAST} \cite{drummond15}---with a birth-death process prior.
This would not enforce the infinite-type assumption, so a set of identical sequences could be placed in disjoint subtrees.
However, such an approach will not scale well with many identical sequences: trees that only differ by exchange of identical sequences will create islands of constant posterior in tree space.
Methods do not currently exist for tree space traversal that avoids moves within such islands.
Even if such methods existed, they would need to be combined with algorithms to infer trees with sampled ancestors \cite{Gavryushkina2014-qp,Gavryushkina2015-ew} as well as multifurcations \cite{Lewis2005-ez,Lewis2015-kv}; even just this combination is not currently available.

Although our methods can be applied to other sequence-based optimality functions besides parsimony, it is important to recognize that \texttt{GCtree} (and indeed any tree inference procedure that deduplicates repeated sequences) contains an inherent weak parsimony assumption: that each unique genotype arose from mutation just once in the lineage and therefore corresponds to a single subtree in the lineage tree, and thus a single node in the GCtree.
Thus it is important to continue to assess the impact of this weak parsimony assumption with simulation that does not make this assumption, as done here.

The \texttt{GCtree} framework can also be extended to non-neutral models.
For example, one could consider a model in which each genotype obtains a random fitness encoded by branching process parameters $\boldsymbol\theta$ that are fixed within a given genotype but randomly drawn by the genotype founder cell upon mutation from its parent.
This will likely necessitate modeling genotype birth time explicitly, rather than restricting to extinct subcritical processes, since a genotype with small abundance may be a result of low fitness or just young age.
One might also consider extending the offspring distribution to separately model synonymous and nonsynonymous mutations.
Synonymous mutations do not change fitness, while nonsynonymous mutations change fitness as described above.
Another direction of extension is to incorporate mutation models specialized to the case of BCR evolution, such as the S5F model \cite{Yaari2013-uw} used in our simulation study.

\section*{Supplementary Material}
Supplementary Table~S1 is available online.

\section*{Acknowledgments}

Arman Bilge originally suggested using branching processes to investigate this process.
This work was greatly improved by discussions with Amaury Lambert and Kristian Davidsen.
The research of WSD was supported in part by a NIH National Human Genome Research Institute Genome Training Grant (5T32HG000035-23).
LM and GDV acknowledge research support from NIH grant R01 AI119006.
VNM and FAM acknowledge research support from NIH grant R01 GM113246 and NSF grants CISE-1564137 and CISE-1561334.
The research of FAM was supported in part by a Faculty Scholar grant from the Howard Hughes Medical Institute and the Simons Foundation, as well as by NIH grant R01 AI120961 (PI Overbaugh).

\section*{Author Contributions}
WSD, VNM, and FAM conceived and developed statistical methods, analyzed the data, and wrote the manuscript.
WSD wrote the software with consultation from FAM.
LM and GDV developed and performed brainbow mouse experiments, developed the intuition formalized by the GCtree algorithm, and consulted on data analysis.
\section*{Competing Interests}
The authors declare that they have no competing financial interests.

\bibliographystyle{natbib}
\bibliography{gctree}

\onecolumn

\section*{Materials and Methods}

\subsection*{\bf An empirical Bayes framework for incorporating genotype abundance in phylogenetic optimality.}

Here we more fully develop the empirical Bayes perspective on our estimator for the model depicted in Figure~\ref{fig:model}a.
This graphical model implies the factorization
\begin{equation}
\mathbb{P}\left(\mathbf{G}, \mathbf{A}, \mathbf{T}, \boldsymbol\theta\right) = \mathbb{P}\left(\mathbf{G}\mid \mathbf{T}\right)\mathbb{P}\left(\mathbf{A}, \mathbf{T}\mid\boldsymbol\theta\right)\mathbb{P}\left(\boldsymbol\theta\right).
\label{condition}
\end{equation}
A hierarchical Bayes treatment would assign a prior $\mathbb{P}\left(\boldsymbol\theta\right)$ (such as uniform over the unit square for the model $\boldsymbol\theta=(p,q)$) and compute the posterior over trees conditioned on the data, marginalizing over $\boldsymbol\theta$:
\begin{align*}
\mathbb{P}\left(\mathbf{T}\mid \mathbf{G}, \mathbf{A}\right) &= \int d\boldsymbol\theta \ \mathbb{P}\left(\mathbf{T}, \boldsymbol\theta\mid \mathbf{G}, \mathbf{A}\right)\\
&= \int d\boldsymbol\theta \ \frac{\mathbb{P}\left(\mathbf{G}, \mathbf{A}, \mathbf{T}, \boldsymbol\theta\right)}{\mathbb{P}\left(\mathbf{G}, \mathbf{A}\right)}\\
&\propto \mathbb{P}\left(\mathbf{G}\mid \mathbf{T}\right)\int d\boldsymbol\theta \ \mathbb{P}\left(\mathbf{A},\mathbf{T}\mid\boldsymbol\theta\right) \mathbb{P}\left(\boldsymbol\theta\right).
\end{align*}
Rather then attempting this integral over $\mathbb{P}\left(\mathbf{A}, \mathbf{T}\mid\boldsymbol\theta\right)$, each evaluation of which requires dynamic programming, we first seek a maximum likelihood estimate for $\boldsymbol\theta$ marginalizing $\mathbf{T}$:
\begin{align}
\hat{\boldsymbol{\theta}} &= \argmax_{\boldsymbol{\theta}} \mathbb{P}\left(\mathbf{G}, \mathbf{A}\mid\boldsymbol\theta\right)\nonumber\\
&= \argmax_{\boldsymbol{\theta}} \sum_\mathbf{T}\mathbb{P}\left(\mathbf{G},\mathbf{A},\mathbf{T}\mid\boldsymbol\theta\right)\nonumber\\
&= \argmax_{\boldsymbol{\theta}} \sum_\mathbf{T}\mathbb{P}\left(\mathbf{G}\mid \mathbf{T}\right)\mathbb{P}\left(\mathbf{A},\mathbf{T}\mid\boldsymbol\theta\right).
\label{eqn:treesum}
\end{align}
Using this point estimate, an approximate posterior over trees is
\begin{equation}
\mathbb{P}\left(\mathbf{T}\mid \mathbf{G}, \mathbf{A}, \hat{\boldsymbol{\theta}}\right) \propto \mathbb{P}\left(\mathbf{G}\mid \mathbf{T}\right) \mathbb{P}\left(\mathbf{A}, \mathbf{T}\mid\hat{\boldsymbol{\theta}}\right).
\label{eqn:combined}
\end{equation}
This formulation embodies an optimality over trees conditioned on both genotype sequence data $\mathbf{G}$ and genotype abundance data $\mathbf{A}$.
Evaluation of $\hat{\boldsymbol{\theta}}$ with \eqref{eqn:treesum} in general requires summation over the space of all trees consistent with the data.

A simple application of this formalism is to augment parsimony-based tree optimality with abundance data.
Let $\mathcal{T}_\mathbf{G}$ denote the degenerate set of maximally parsimonious trees given $\mathbf{G}$ (each of which has the same total genotype sequence distance over its edges).
Encode parsimony optimality as a $\mathbb{P}\left(\mathbf{G}\mid \mathbf{T}\right)$ assigning uniform weight to each tree in $\mathcal{T}_\mathbf{G}$, and zero elsewhere.
In this case, \eqref{eqn:theta} becomes
\begin{equation}
\hat{\boldsymbol{\theta}} = \argmax_{\boldsymbol{\theta}} \sum_{\mathbf{T}\in\mathcal{T}_\mathbf{G}}\mathbb{P}\left(\mathbf{A}, \mathbf{T}\mid \boldsymbol\theta\right),
\label{eqn:thetaParsimony}
\end{equation}
and \eqref{eqn:combined} becomes
\begin{equation}
\mathbb{P}\left(\mathbf{T}\mid \mathbf{G}, \mathbf{A}, \hat{\boldsymbol{\theta}}\right) \propto
\begin{cases}
\mathbb{P}\left(\mathbf{A}, \mathbf{T}\mid \hat{\boldsymbol{\theta}}\right), &t \in \mathcal{T}_g\\
0, &t \notin \mathcal{T}_g
\end{cases}.
\label{eqn:combined_parsimony}
\end{equation}
With \eqref{eqn:combined_parsimony}, we have a framework using abundance information to distinguish among the otherwise equally optimal trees presented by a parsimony analysis.
In our application, we use a subcritical infinite-type binary Galton-Watson branching process model for the lineage tree, and describe a recursion for computing GCtree likelihoods $\mathbb{P}\left(\mathbf{A}, \mathbf{T}\mid \hat{\boldsymbol{\theta}}\right)$ by dynamic programming to marginalize over compatible lineage trees.

\subsection*{\bf Dynamic programming to marginalize lineage tree structure.}

We derive a recurrence for $f_{a \tau}(p, q) = \mathbb{P}\left(A_i=a,T_i=\tau\mid \boldsymbol\theta=(p, q)\right)$ by marginalizing over the outcome $\{C, M\}$ of the branching event at the root of the lineage subtree for genotype $i$ (the first cell of type $i$).
We will use that $a$ and $\tau$ are the sum over two iid processes for the left and right clonal branches.
We temporarily suppress the parameters $\boldsymbol\theta=(p, q)$, writing $f_{a \tau}$ for notational compactness.
In the case $\{C=2, M=0\}$,
\begin{align}
\mathbb{P}\left(A_i=a,T_i=\tau \mid C=2, M=0\right) &= \sum_{a'=0}^a \sum_{\tau'=0}^\tau f_{a' \tau'} f_{a-a',\tau-\tau'}.
\end{align}
As this is the convolution of $f_{a \tau}$ with itself, we denote it as $f_{a \tau}^{*2}$.
Marginalizing over all outcomes $\{C, M\}$, we have
  \begin{align}
  f_{a \tau} &= \sum_{(c, m) \in \mathbb{N}^2} \mathbb{P}\left(A_i=a,T_i=\tau \mid C=c, M=m\right) \mathbb P\left(C=c, M=m\right) \nonumber\\
  &= \delta_{a 1}\delta_{\tau 0}(1-p) + f^{*2}_{a \tau}p(1-q)^2+(1-\delta_{\tau 0})f_{a, \tau-1}2pq(1-q) + \delta_{a 0}\delta_{\tau 2}pq^2\nonumber\\
  &=
  \begin{cases}
        0 & a=0, \tau=0,1,\\
        (1-p) & a=1, \tau=0, \\
        pq^2 & a=0, \tau=2, \\
        f^{*2}_{a 0}p(1-q)^2& a>1, \tau=0,\\
        f_{a, \tau-1}2pq(1-q) + f^{*2}_{a \tau}p(1-q)^2& \text{otherwise},
  \end{cases}
  \label{eqn:recursion}
  \end{align}
where $\delta_{\cdot \cdot}$ denotes the Kronecker delta function.
In light of the first case, the convolutional square may be written as
\begin{equation*}
f^{*2}_{a \tau} = \sum_{(a',\tau')\notin\{(0,0),(a,\tau)\}}f_{a' \tau'} f_{a-a',\tau-\tau'},
\end{equation*}
showing that there are no terms containing $f_{a \tau}$ on the RHS of \eqref{eqn:recursion}.
The GCtree node likelihood $f_{a \tau}$ is thus amenable to computation by straightforward dynamic programming.

\subsection*{\bf The GCtree likelihood factorizes by genotype.}

We argue that the joint distribution over all nodes in a GCtree factorizes by genotype (Figure~\ref{fig:model}d):
\begin{equation}
\mathbb{P}\left(A_1=a_1, T_1=\tau_1, \dots, A_N=a_N, T_N=\tau_N\right) = \prod_{i=1}^N f_{a_i \tau_i}.
\label{eqn:likelihood2}
\end{equation}

Since $\tau_1$ is the number of children of node 1 (the root node), the children of the root node are indexed in level order by $2,\dots,1+\tau_1$.
Let $\Lambda_i$ denote the set of indices of the nodes of the subtree rooted at node $i$, so $\Lambda_2,\dots,\Lambda_{1+\tau_1}$ refer to sister subtrees rooted on each of the $\tau_1$ children of the root.
Using the definition of conditional probability, and since sister subtrees are independent, we have
\begin{align*}
\mathbb{P}\left(a_1, \tau_1, \dots, a_N, \tau_N\right) &= \mathbb{P}\left(a_2, \tau_2, \dots, a_N, _N\mid a_1, \tau_1\right) \mathbb{P}\left(a_1, \tau_1\right)\\
&= f_{a_1 \tau_1}\prod_{i=1}^{1+\tau_1}\mathbb{P}\left(\{(a_j,\tau_j):j\in\Lambda_i\}\right),
\end{align*}
where random variable notation has been dropped for notational compactness.
Now, within each subtree factor we may reindex in level order (that is, level order in that subtree) starting from 1.
We then pull out factors $f_{a_2 \tau_2},\dots,f_{a_{1+\tau_1} \tau_{1+\tau_1}}$ corresponding to the root nodes of the sister subtrees (children of the original root).
We obtain \eqref{eqn:likelihood2} by applying this logic recursively.
Restoring the offspring distribution parameters, we recognize this as the distribution needed in \eqref{eqn:MLE} and \eqref{eqn:theta} to rank trees in a parsimony forest:
\begin{equation}
\mathbb{P}\left(\mathbf{T}=(\tau_1,\dots,\tau_N), \mathbf{A}=(a_1,\dots,a_N)\mid\boldsymbol\theta=(p, q)\right)
= \prod_{i=1}^N f_{a_i \tau_i}\left(p, q\right),
\label{eqn:likelihood3}
\end{equation}
where $f_{a_i \tau_i}\left(p, q\right)$ is computed by dynamic programming using \eqref{eqn:recursion}.

\begin{figure}[p]
\centering
\includegraphics[width=\linewidth]{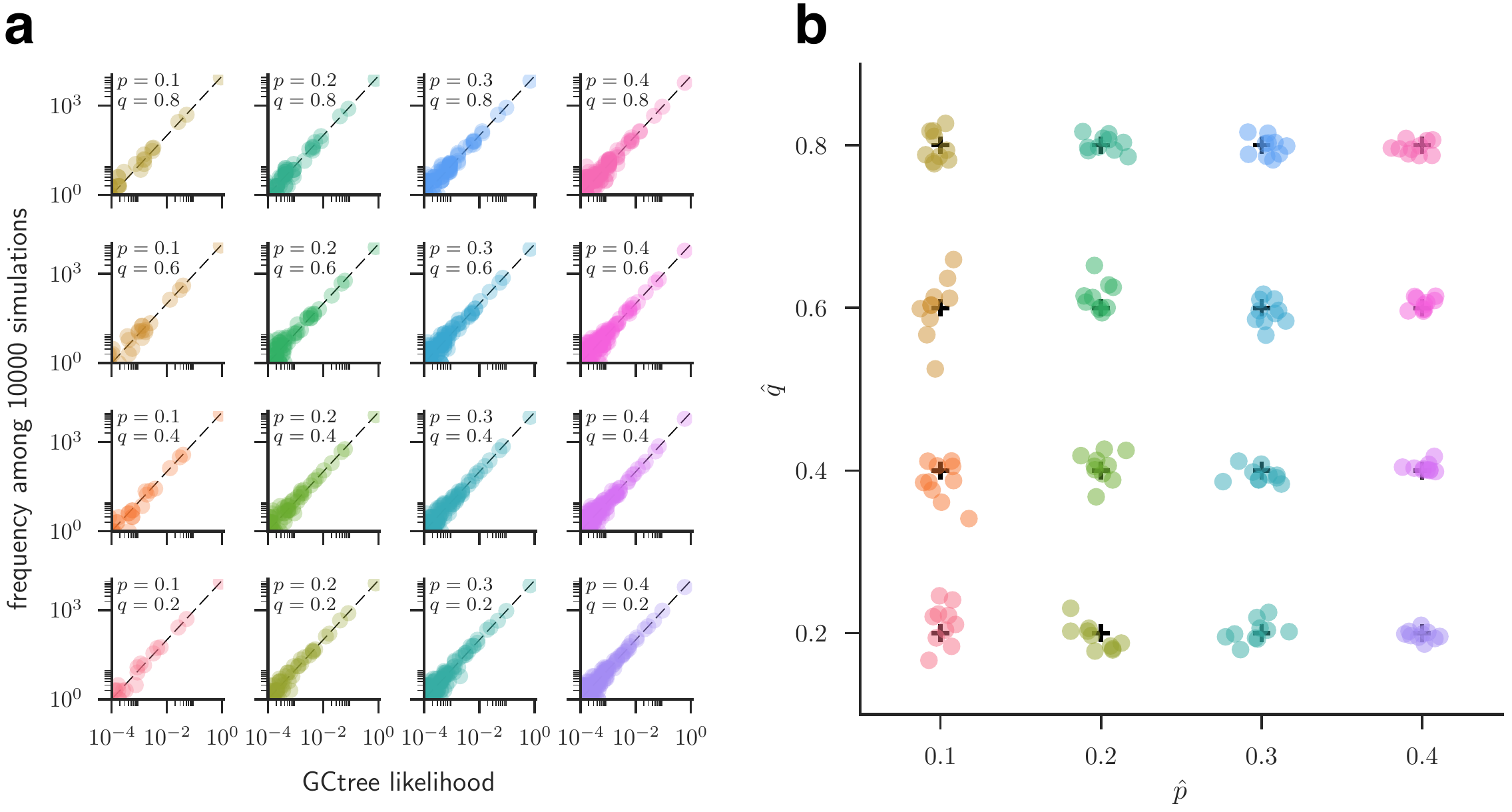}
 \vspace{-10pt}
\caption{
Numerical validation of GCtree likelihood.
Colors indicate simulation parameters.
\textbf{(a.)} At each parameter value ($p$, $q$), 10,000 Galton Watson processes were simulated.
For each distinct GCtree, the likelihood was computed according to \eqref{eqn:likelihood3}, and the frequency of the tree (number of times this distinct tree occurs among the 10,000) was recorded. Dashed lines indicate the expected frequencies (likelihood multiplied by 10,000).
\textbf{(b.)} Each set of 10,000 trees was partitioned into 10 groups of 1000, and maximum likelihood estimates ($\hat p$, $\hat q$) were computed for each set of 1000 by numerical maximization of \eqref{eqn:likelihood3}.}
\label{fig:likval}
\end{figure}

Numerical validation of the GCtree likelihood is summarized in Figure~\ref{fig:likval} using 10,000 Galton-Watson process simulations at each of several parameter values.
The likelihood accurately recapitulates tree frequencies, and simulation parameters are recoverable by numerical maximum likelihood estimation.

\subsection*{\bf Simulation details.}
To provide for a more challenging \textit{in silico} validation study, several biological realisms were built into our simulation that defied simplifying assumptions in the \texttt{GCtree} inference methodology.

\subsubsection*{Arbitrary offspring distribution.}
The recursion \eqref{eqn:recursion} used to compute GCtree likelihood components specifies a binary branching process, and such an approach would in general require an offspring distribution with bounded support on the natural numbers.
Our simulation implements an arbitrary offspring distribution with no explicit bounding.
In the results that follow, we used a Poisson distribution with parameter $\lambda$ for the expected number of offspring of each node in the lineage tree.

\subsubsection*{Context sensitive mutation.}
To generate mutant offspring, all offspring sequences (drawn from a Poisson as described above) were subjected to a sequence-dependent mutation process.
The SHM process is known to introduce mutations in a sequence context-dependent manner, with certain hot-spot and cold-spot motifs \cite{Dunn-Walters2360, Spencer5170}.
We used a previously published 5-mer context model S5F \cite{Yaari2013-uw} to compute the mutabilities $\mu_1$, $\dots$, $\mu_\ell$ of each position $1$, $\dots$, $\ell$ within a sequence of length $\ell$ based on its local 5-mer context.
This model also provided substitution preferences among alternative bases given the 5-mer context.
To compute mutabilities for beginning and ending positions without a complete 5-mer context, we averaged over missing sequence context.

Although existing code can simulate a mutational process parameterized by S5F on branches of a fixed tree with a pre-specified number of mutations on each branch \cite{gupta2015}, in our simulations we wanted the number of mutations on the branches to be determined by the sequence mutability as it changes via mutation across the tree.
For example, as an initial mutation hotspot motif acquires mutations down the tree, its mutability typically degrades as it diverges from the original motif.
We defined the mutability of the sequence as a whole by the average over its positions $\mu_0=\frac{1}{\ell}\sum_{i=1}^\ell \mu_i$.
We defined a baseline mutation expectation parameter $\lambda_0$ as a simulation parameter, and the number of mutations $m$ any given offspring sequence received was drawn from a Poisson distribution.
The Poisson parameter was modulated by the sequence's mutability $m\sim\text{Pois}(\mu_0 \lambda_0)$, so that more mutable sequences tended to receive more mutations.
Given $m>0$, the positions in the sequence to apply mutations were chosen sequentially as follows.
A site $j$ to apply the first mutation was drawn from a categorical distribution using the site-wise mutabilities to define relative probability of choosing each site $j\sim\text{Cat}(\mu_1,\dots,\mu_\ell)$.
We mutated the site using a categorical distribution over the three alternative bases parameterized by the substitution preferences defined by the site's context.
We then updated mutabilities $\mu_0$ and $\mu_1$, $\dots$, $\mu_\ell$ as necessary to account for contexts that had been altered by the mutation.
This process was repeated $m$ times.

Since the mutability of each node in the lineage tree will depend on the mutation outcome of its parent, the GCtree likelihood components will not factorize by genotype.
Because the probability of mutation is sequence-dependent, the topology of the GCtree will be sequence-dependent.
Therefore, the generative assumptions of the empirical Bayes inference do not hold in this simulation scheme, nor does the offspring distribution equivalence across lineage tree nodes specified by \eqref{eqn:offspring}.

\subsubsection*{Sampling time.}
Our inference model specifies a subcritical branching process run until extinction, and sampling of all terminated nodes (leaves).
Our simulation more realistically assigns a discrete time of sampling parameter $t$ (number of time steps from root), and thus does not need to constrain the offspring distribution to achieve subcriticality.
At the specified time, extant nodes can be sampled, so all genotypes that terminated or mutated at a prior times are not observed.
Alternatively, a parameter $N$ specifying the desired number of simulated observed sequences may be passed, in which case the simulation runs until a time such that at least $N$ sequences exist (unless terminated).
Genotypes born at different times will be sampled under a process with different effective sampling times since birth.
Thus this sampling time parameter also increases dependence between genotypes, further distancing the simulation model from the inferential model.

\subsubsection*{Incomplete sampling.}
We introduce imperfect sampling efficiency with a parameter $n$ for the number of simulated sequences that end up in the simulated sample data (\texttt{FASTA}), requiring $n \leq N$.
This violates the inferential assumption of complete sampling, and renders the true genotype abundances latent variables (which a more complete likelihood approach might aim to marginalize out).

\subsubsection*{Repeated genotypes.}
Our simulation is seeded with an initial naive BCR sequence, from which randomly mutated offspring are created.
Because there is no built-in restriction that the same sequence cannot arise along different branches (or mutations could be reversed), the model assumption of infinite types---such that identical sequences can be associated with a single genotype subtree---does not necessarily hold.
When this assumption is violated the tree must necessarily be incorrect.

\subsection*{\bf Calibrating simulation parameters using summary statistics.}

We defined several summary statistics on sequences equipped with abundances which were used to calibrate simulation parameters representative of a regime similar to experimental data.
We chose these statistics to reflect information relevant to tree inference, but not actually require tree inference, so as to avoid circularity.
Denote $g_0\in \mathbf{G}$ as the naive BCR (root genotype) and $d_H(\cdot, \cdot)$ as the Hamming distance function between two sequences.
Given simulation or experimental data $\mathbf{G}$ and $\mathbf{A}$, we characterize the degree of mutation (from naive BCR) in the lineage by the set of Hamming distances of the observed genotypes from the naive genotype: $\{d_H(g, g_o), g\in \mathbf{G}\}$.
For a given genotype $g_i\in \mathbf{G}$, we can compute its number of Hamming neighbors in the data $\eta_i=|\{g_j\in \mathbf{G}:d_H(g_i, g_j)=1\}|$.

\begin{figure}[h]
\centering
\includegraphics[width=\linewidth]{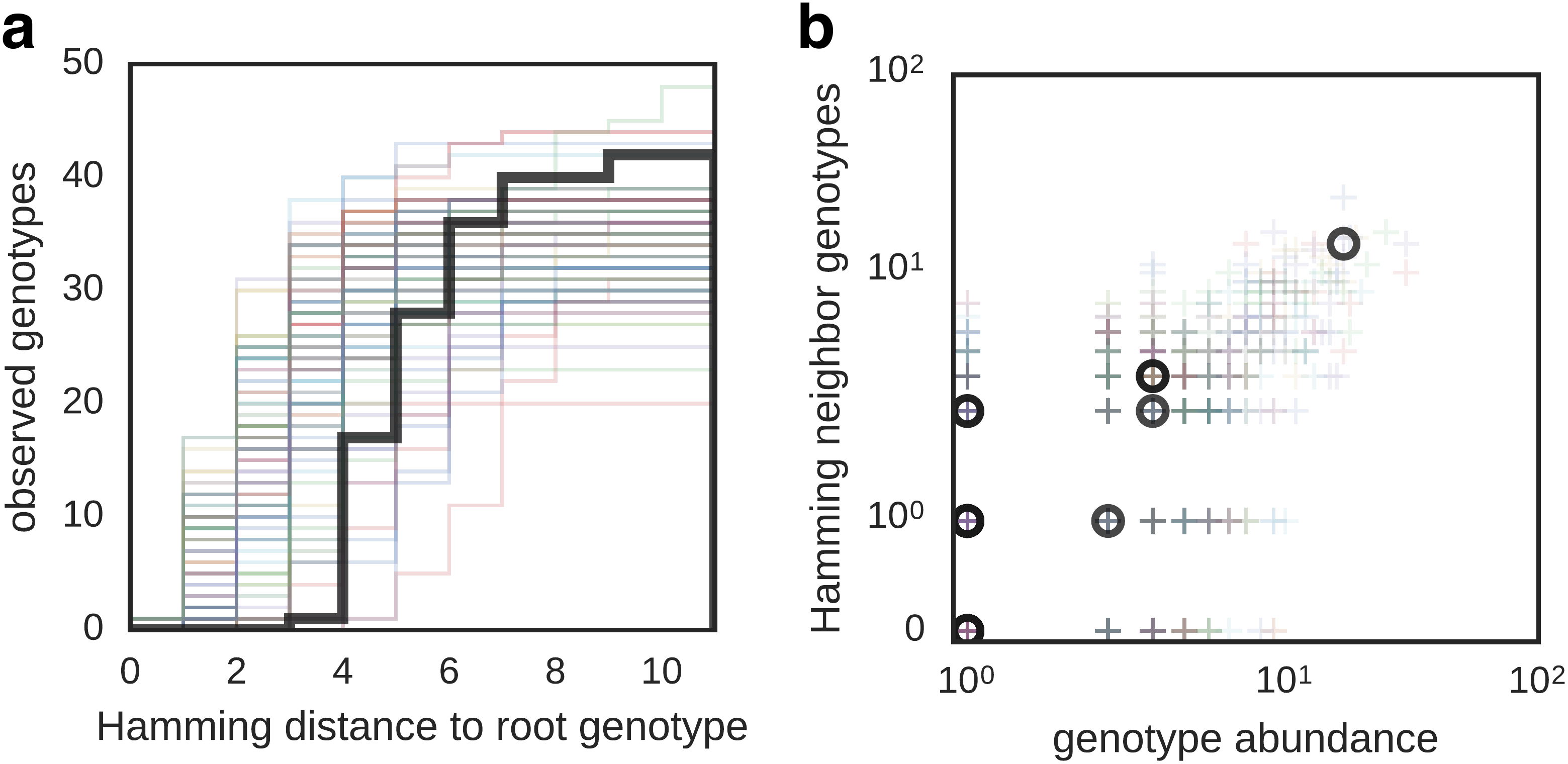}
 \vspace{-10pt}
\caption{
Simulation summary statistics.
simulation parameters: $\lambda=1.5$, $\lambda_0=.25$, $N=100$, $n=65$.
\textbf{(a.)} The empirical CDF over genotypes of Hamming distance to the naive genotype for 100 simulations (colors) and germinal center BCR data (black).
\textbf{(b.)} The distribution over genotypes of number of Hamming neighbors and genotype abundance for 100 simulations (colors) and germinal center BCR data (black).}
\label{fig:simstat}
\end{figure}

A simulation is specified by parameters $\left(\lambda, \lambda_0, N\text{(or $t$)}, n\right)$, a mutability model (here S5F \cite{Yaari2013-uw}), and an initial sequence.
We found parameters $\left(\lambda=1.5, \lambda_0=0.25, N=100, n=65\right)$ produced simulations that were comparable to experimental data under these statistics.
The experimental data used for comparison, consisting of 65 total BCR V gene sequences from a single germinal center lineage, is described in the following section.
Figure~\ref{fig:simstat} depicts these summary statistics for 100 simulations, compared to experimental BCR data.

\subsection*{\bf Germinal center BCR sequencing.}

Germinal center B cell lineage tracing and B cell receptor sequencing was performed as previously described \cite{Tas2016-lq}.
Full length IgH and IgL sequences from lymph node 2 germinal centers 1 and 2 from this reference were used for empirical validation results, while V gene sequences only (which are not dependent on \texttt{partis}-inferred naive sequences) were used for calibrating simulation parameters.

\subsection*{\bf Bootstrap support.}

We computed bootstrap support values for edges on a GCtree extending the standard approach \cite{felsenstein1985confidence}: we resampled columns from the alignment $G$ 100 times with replacement, generating an inferred GCtree (maximum GCtree likelihood among equally parsimonious trees) for each.
Each edge is equivalent to a bipartition of observed genotypes obtained by cutting the edge; such a bipartition is typically referred to as a \emph{split}.
We compute the number of bootstrapped trees that contain the same split, and annotate the edge with this number.
Because resampling the alignment $G$ can produce repeated genotypes, there can exist ambiguity about how to perform genotype collapse of a parsimony tree.
We simply group genotypes in the bootstrap analysis that collapse to identical genotypes.
For example, if two observed sister genotypes with resampled sequences are both identical in sequence to their mutual parent, both have a claim on collapsing into the parent.
When collapsing this tree, both genotypes will be associated with this collapsed node, rather then just a single one.

\subsection*{\bf Data availability.}

Germinal center BCR sequence data can be found in Supplementary Database S1 of Tas et al. \cite{Tas2016-lq}, lymph node 2 and germinal center 1.

\subsection*{\bf Software availability.}

The \texttt{GCtree} source code is available at \url{github.com/matsengrp/gctree} and accepts sequence alignments in \texttt{FASTA} or \texttt{PHYLIP} format as input.
It is open-source software under the GPL v3.

\beginsupplement

\begin{landscape}

\begin{figure}[p]
\centering
\includegraphics[width=\linewidth]{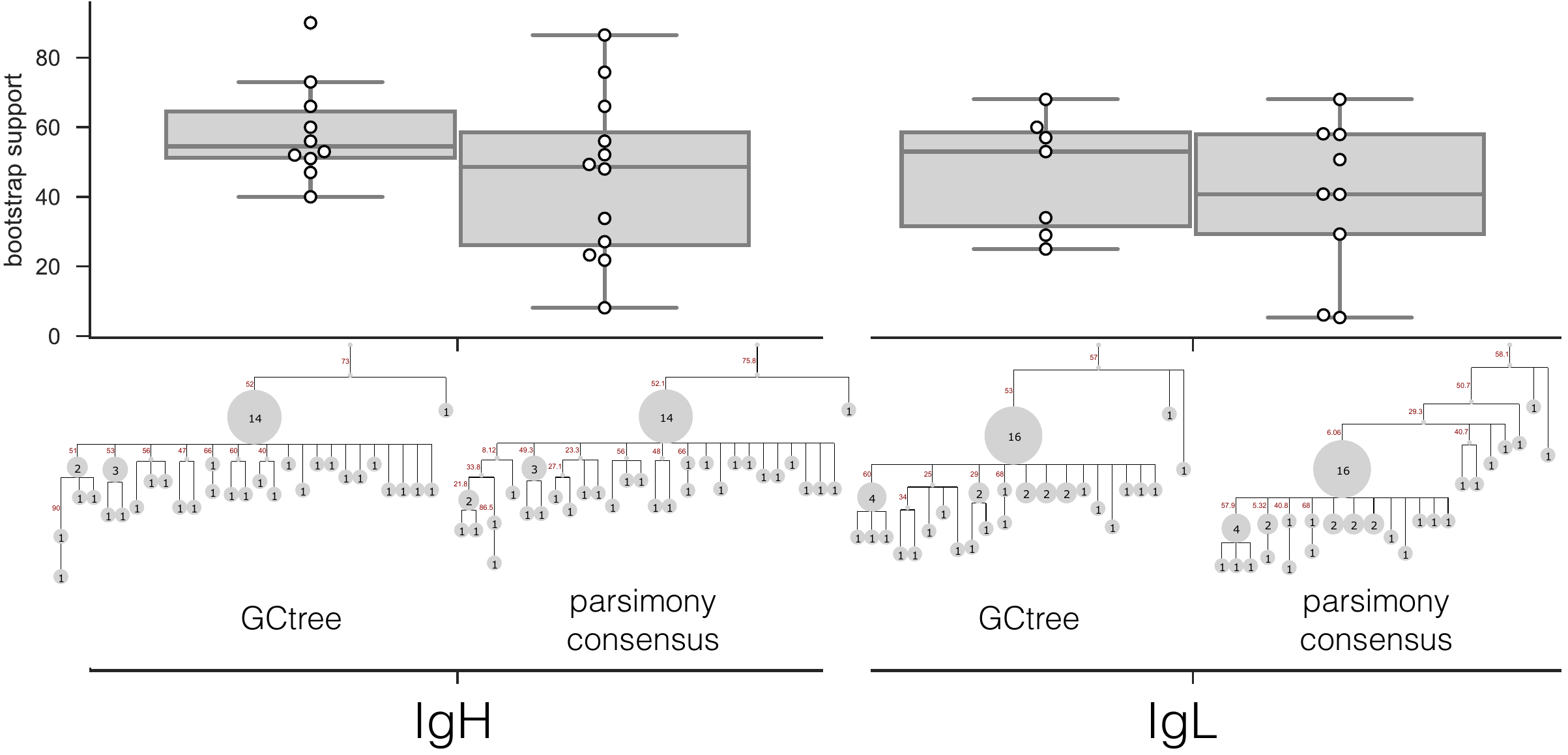}
 \vspace{-10pt}
\caption{
Bootstrap support comparison.
Support values among 100 bootstrap samples are shown for splits in the \texttt{GCtree} result and consensus parsimony tree for IgH and IgL sequence data from the same germinal center lineage.
\label{fig:support}}
\end{figure}

\begin{figure}[p]
\centering
\includegraphics[width=\linewidth]{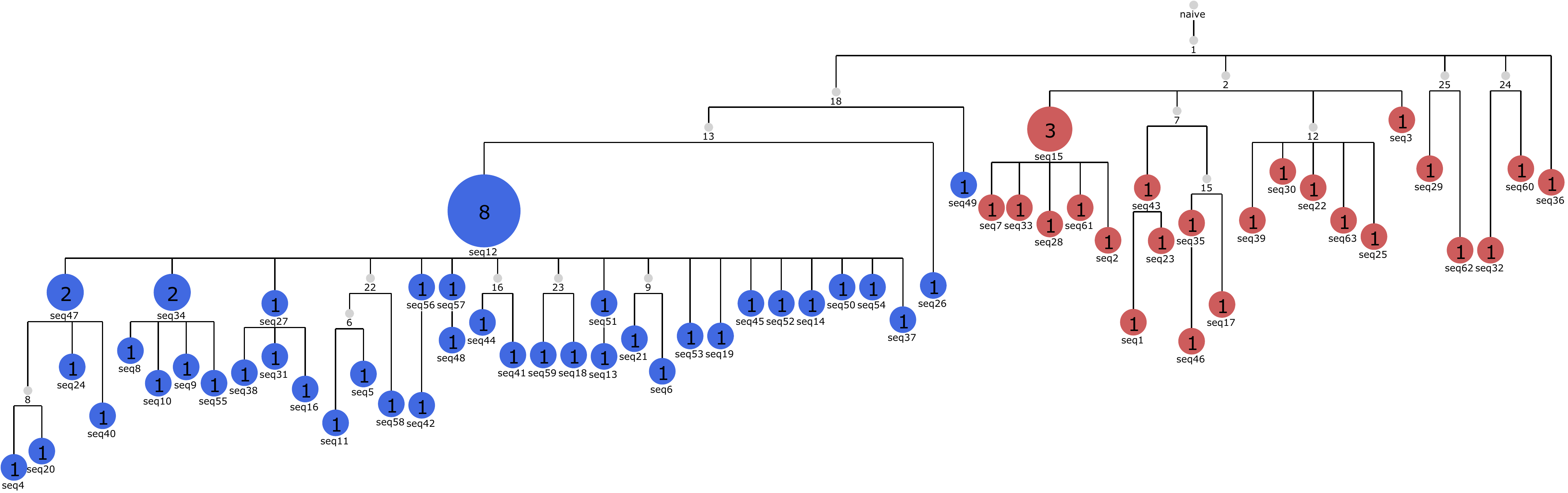}
 \vspace{-10pt}
\caption{
Version of Figure~\ref{fig:empval}a with nodes annotated below by sequence names from the Supplementary \texttt{FASTA} alignment file.
In this alignment file, columns 1--303 represent IgH sequences and 304--546 represent IgL sequences.
\label{fig:labelled_tree}}
\end{figure}

\end{landscape}

\end{document}